\documentclass[%
 aip, numerical,
rsi,%
 amsmath,amssymb,
 reprint,%
]{revtex4-2}

\usepackage[utf8]{inputenc}
\usepackage{graphicx}
\usepackage{dcolumn}
\usepackage{bm}
\usepackage{amsfonts}
\usepackage{tabularx, booktabs}
\usepackage{siunitx}
\usepackage{array}

\begin{document}
\preprint{AIP/123-QED}

\title{Experimental observation of trajectories beyond the long in high order harmonic generation}
 
 \author{S. Bengtsson}
 \thanks{Authors contributed equally to the paper}
 \affiliation{ 
Department of Physics, Lund University, Box 118, SE-222 10 Lund, Sweden.
}%

\author{E. R. Simpson}
 \thanks{Authors contributed equally to the paper}
\affiliation{ 
Department of Physics, Lund University, Box 118, SE-222 10 Lund, Sweden.
}%

\author{N. Ibrakovic}

\affiliation{ 
Department of Physics, Lund University, Box 118, SE-222 10 Lund, Sweden.
}%

\author{S. Ek}%
 
\affiliation{ 
Department of Physics, Lund University, Box 118, SE-222 10 Lund, Sweden.
}%

\author{A. Olofsson}
 
 \affiliation{ 
Department of Physics, Lund University, Box 118, SE-222 10 Lund, Sweden.
}%

\author{T. Causer}
 
\affiliation{ 
Department of Physics, Lund University, Box 118, SE-222 10 Lund, Sweden.
}%

\author{J. Mauritsson}
\email{johan.mauritsson@fysik.lth.se}
\affiliation{ 
Department of Physics, Lund University, Box 118, SE-222 10 Lund, Sweden.
}%

\date{\today}

\begin{abstract}

We experimentally observe longer than long trajectory influence in high order harmonic generation (HHG) by varying the peak intensity of the driving laser field through either direct attenuation, or by chirping the laser pulse. Using a theoretical Gaussian beam model to simulate spatial interference resulting from quantum path interference we show that the measured interference patterns cannot be solely explained by the well established short and long trajectories. The structure change is most prominent for the more divergent, off-axis components of the lower plateau harmonic region, affecting the direction and amplitude of the extreme ultraviolet light emitted, and is thus of importance for understanding and controlling the fundamentals of the HHG process.

\end{abstract}

                              
\maketitle

\section{\label{sec:level1} Introduction}

\begin{figure}[!ht]
    \centering
    \includegraphics{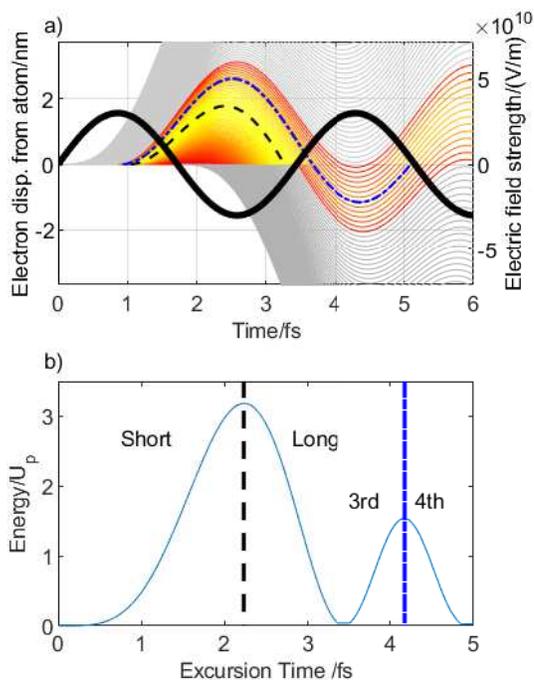}
    \caption{a) Classically calculated electron trajectories emitted for one half-cycle of the sinusoidal driving laser field (thick black line). The recombining trajectories are shown in color with the variation towards yellow representing an increased kinetic energy at return. b) Kinetic energy of the returning electrons as a function of excursion time for the electron in the continuum. Four groups of trajectories are defined. The dashed black and dot-dash blue trajectories in a) and b) correspond to the boundaries between short-long and third-fourth groups respectively.}
    \label{fig:traj}
\end{figure}

High-order harmonic generation (HHG) \cite{McPhersonJOSAB1987, FerrayJPB1988} based extreme ultraviolet (XUV) light sources present an excellent tool to probe and control ultrafast electron dynamics in atoms \cite{GisselbrechtPRL1999, UiberackerNature2007,GoulielmakisNature2010}, molecules \cite{SorensenJCP2000, BauerPRL2001, Gagnonscience2007, CalegariSci2014} and solids \cite{CavalieriNature2007}.
The technique of HHG has been studied and exploited since the late 1980s, pushing the scientific boundary for temporal studies \cite{PaulScience2001, HentschelNature2001, CorkumNP2007}, and is now making its way into industry. Despite being a well established technique for generating coherent, sub-femtosecond pulses of XUV light, there are still questions and unknowns within the fundamental HHG process warranting further study and explanation. An established simple semi-classical model \cite{KrausePRL1992, CorkumPRL1993, LewensteinPRA1994}, that can well describe the qualitative results of this process has generally been used by the community to increase our understanding. In this model the generated photons are the result of tunnel ionization, propagation and recombination of electrons bound to the atom in a strong laser field. The propagation of the ionised electron wave packet in the intense laser field can be treated as a series of classical charged free bodies moving in a varying electric field, greatly simplifying the calculation complexity of the process. Each possible path that the electron can take depends on the phase of the field when it is ionized, and is referred to as an electron trajectory, see Figure \ref{fig:traj}.

In this simple picture the different trajectories each result in a specific XUV photon energy generated upon recombination. The different recombining trajectories can either return directly, or after passing the core at least one time. Both the current and historical focus from the community has centered on the electron trajectories that return directly, corresponding to those with the shortest electron paths. One reason for this is that the main features of the HHG spectrum can be well explained by just these trajectories. 
This raises the question of whether the longer trajectories that pass the core before recombination contribute at all to the HHG signal, and if so, in what way? It has been predicted theoretically that some of the longer trajectories should contribute significantly to the generated XUV signal \cite{GaardeOE2001, TatePRL2007,MillerPRA2014, LiSREP2016}, and experimentally interference between the short and long trajectories has been observed \cite{ZairPRL2008}. Contributions from other trajectories, however, have not yet been identified experimentally.

In this article we present experimental results showing the evolution of the harmonic signal as the peak intensity of the driving laser pulse is varied, and find results that can be explained through the inclusion of longer trajectory contributions. The peak intensity was varied using two methods, both through direct attenuation of the laser pulse energy, as well as through temporal chirping of the pulse, as per previous studies \cite{CarlstromNJP2016, CsizmadiaNJP2021}. When comparing our results from either low or high harmonic orders we found two distinct experimental signatures occurring. Using a simple theoretical model we could identify these signatures as resulting from either long-long or long-longer trajectory interference.

This article is organised in the following way: we begin by discussing the detected quantum path interference (QPI) that arises from the different trajectories, and introduce the simple Gaussian model used to explain our experimental observations. Following this, we present the experimental setup and methods used to obtain and process the recorded data. Next we compare the direct attenuation and chirp methods for reducing the peak intensity of the driving laser pulse. Finally, we draw conclusions regarding the longer trajectories based on the QPI plots and discuss their influence on the HHG signal.

\section{Quantum Path Interference}

High order harmonic generation is produced by focusing a powerful laser pulse into a target medium. This strongly non-linear process requires high peak intensities to distort the Coulomb potential for the bound electrons, enabling tunnel ionization of part of the bound electron wave packet. The ionised electron wave packet is then accelerated in the oscillating driving laser field, evolving as it travels. In the event of return to the origin, interference occurs with the portion of the remaining bound electron wave packet resulting in an oscillating dipole and emitted light. This quantum mechanical process produces the surprising, but well known, shape of a high order harmonic spectrum, where a long plateau region of similar yield is produced for the emitted XUV spectrum, before a sharp cut-off region. To simplify the discussion, we will initially focus on the result from just a single atom before considering the effects from many atoms.

  \begin{figure}
    \centering
    \includegraphics{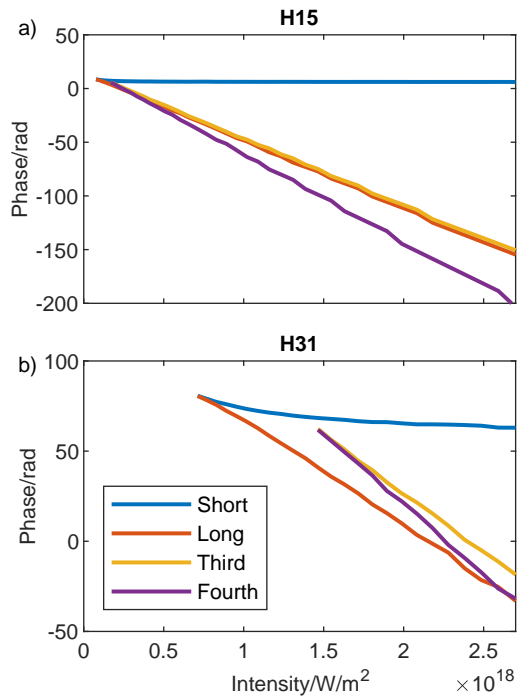}
    \caption{Classically calculated phase variation as a function of peak intensity of the driving laser field for two different harmonic energies generated in argon using a driving wavelength of 1030\,nm. a) Harmonic 15 (18.1 eV), b) Harmonic 31 (37.3 eV).}
    \label{fig:phase_intensity}
\end{figure}

Approximating the quantum nature of the HHG process into a semi-classical picture allows for a more intuitive understanding. In this semi-classical picture the spatial paths travelled by the electron wave packet, also called quantum paths, are modelled as trajectories for free electrons emitted at different phases of the driving laser field. For the case of a symmetric, sinusoidal field shape, the same spectrum is produced for each half-cycle of the laser field, resulting from symmetric trajectories. By focusing on the trajectories originating within a single half-cycle of this laser field, there will be a number of trajectories that return to the core, as well as non-returning trajectories, see fig \ref{fig:traj}\,a). Assuming that the free electrons start from rest, upon return to the core they have a range of different kinetic energies starting from zero. The peak kinetic energy value depends on the peak intensity, waveform and wavelength of the driving laser field.

As seen in fig. \ref{fig:traj}\,a), classical trajectories that return to the core can have passed the core multiple times before recombination. Sorting the trajectories originating from a single half-cycle according to their return time, the trajectories can be grouped according to their properties, such that the first returning group, the short trajectories, have increasing kinetic energy with increasing excursion time upon return to the core, fig. \ref{fig:traj}\,b). The next group to return, the long trajectories, then have decreasing kinetic energies with increasing excursion time. The following groups, called the third, fourth, and so on, repeat this pattern. It should be noted that the trajectory groups longer than long follow trajectories that could have been categorised as long trajectories if recombination had occurred directly. The simple picture of a charged free particle oscillating in the driving laser field means that no interaction with the Coulomb potential is included. This is especially relevant for the trajectories passing the core before recombination as the phase will be affected by the re-scattering event \cite{LiSREP2016}. In fig. \ref{fig:traj}\,a) and b) the cut-off trajectories between short-long and third-fourth are highlighted using dashed and dot-dashed lines respectively. The difference in returning kinetic energy between the two cut-off trajectories can clearly be seen in fig. \ref{fig:traj}\,b) \cite{LewensteinPRA1994}.

 The different kinetic energies of the returning electrons result in a continuum of XUV frequencies with different phases. In each trajectory group there is a unique path producing each XUV frequency. For a particular emitted XUV frequency there are therefore multiple possible trajectories, coming from different trajectory groups, that can be followed to produce this frequency. The emitted photons resulting from these different trajectory paths will, however, not have the same phase due to the different driving laser field conditions experienced before recombination.

In fig. \ref{fig:phase_intensity} , we see the classically calculated phase variation, for harmonics 15 and 31, as a function of driving field intensity. Not only do we see that there is an approximately linear relationship between the phase of the emitted frequency and the intensity of the driving laser field, coinciding with previous studies \cite{VarjuJMO2005}, but we can also observe a number of key features that help us to understand and explain our experimental observations. 
 The first is that for higher photon frequencies we observe an increasing separation between the cut-off intensities for the short-long and third-fourth trajectory groups, arising due to the influence of the ionisation potential of the atom. These cut-off intensities can be seen from fig. \ref{fig:phase_intensity} in the crossing points, which mark the singular point between adjacent trajectory groups where the phase is the same. As shown explicitly in fig. \ref{fig:phase_intensity}\,b) and fig. \ref{fig:traj}\,b), in practice higher photon frequencies can have returning short and long trajectories without contributions from third and fourth. The second key-feature is that the third trajectory phase-gradient is lower than that for the long trajectory for low frequencies, fig. \ref{fig:phase_intensity}\,a), resulting from second order dispersion. As we look at higher frequencies, the relative phase-gradient changes so that the third trajectory gradient is instead higher than the long trajectory gradient.

  \begin{figure}[!ht]
    \centering
    \includegraphics[width=0.42\textwidth]{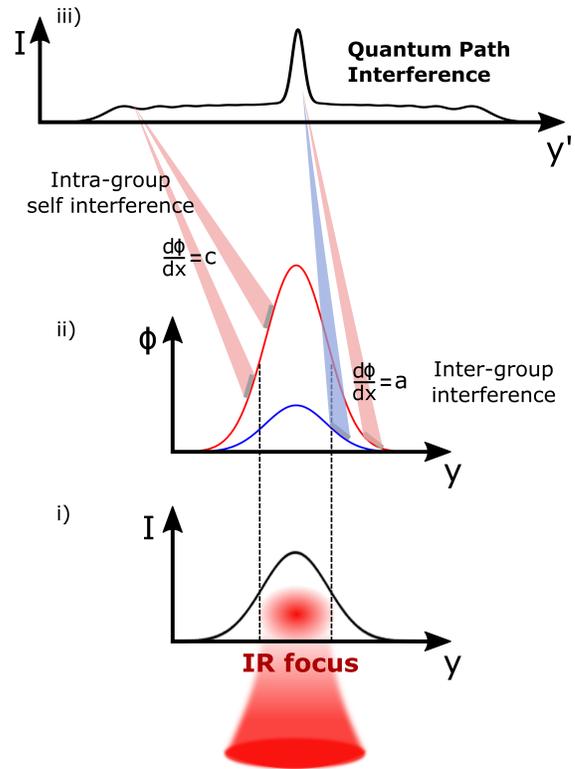}
    \caption{Schematic diagram illustrating how the spatial intensity profile within the focal volume of the laser-gas interaction (i) results in quantum path interference at the detector (iii). With the approximation in eq. \ref{eq:1} the intensity profile (i) is proportional to the spatial phase profile (ii) with different $\alpha$-values for the different trajectory groups. Interference can occur both from within a single trajectory group, e.g. long-long (intra-group) interference, as well as between different trajectory groups, e.g. short-long (inter-group) interference (ii).}
    \label{fig:macroscopic_interference}
\end{figure}

Using the result in fig. \ref{fig:phase_intensity} we can to the simplest approximation make a linear model between the phase and intensity.  Formulating this approximate linearity between phase $\Phi^q _{i}$ and intensity $I$, in the commonly used way, we have the following equation:
 
 \begin{equation}
    \Phi^q _{i} = \alpha_{i}I + \phi_{i}
    \label{eq:1}
\end{equation}
where $\alpha_{i}$ is known simply as the alpha parameter for the different $i^{th}$ trajectory groups, and  $\phi_{i}$ is a constant offset. As seen in fig. \ref{fig:phase_intensity} the gradient of the different fitted lines for the different trajectory groups, the alpha parameters, gives us information about how strongly the intensity effects the phase for each group. More elaborate models also exist that include further effects \cite{GuoJPBAMOP2018}, especially near the cut-off regions. For our qualitative understanding the linear model is sufficient.

So far, this discussion has considered the result from just a single atom.  In the reality of an experiment, there is a focal spot volume containing many atoms with a corresponding intensity distribution. This will lead to a different peak intensity of the driving laser field being felt by the different atoms within this volume. For a Gaussian laser beam profile, the focal spot will also be Gaussian, leading to a reduced peak intensity experienced by an atom towards the wing of the focus compared with one at the center. Due to the approximated linear relation between phase and intensity as described in eq.\ref{eq:1}, this translates into a varying phase across the focal spot profile for each emitted frequency for each trajectory group. When considering the effects from the multiple contributing atoms across the focal profile, this results in Gaussian wavefronts of the emitted frequency.

 \begin{figure}
     \centering
     \includegraphics[width=0.43\textwidth]{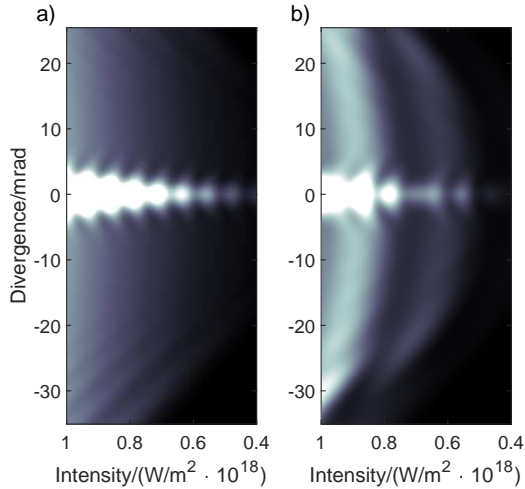}
     \caption{Simulated QPI plots using the simple Gaussian beam model to show the influence of a longer trajectory. In a) only short and long trajectories are present. In b) the third trajectory group is also included. This shows the two different signatures in the off-axis part of the QPI plot: the linear stripes signature of the long-long intra-trajectory interference in a), and in b) the curved signature arising from interference between trajectory groups (inter-group interference). The parameter values used for the simulation can be found in Table\,\ref{table:1} for Harmonic 15.}
     \label{fig:sim_qpi}
 \end{figure}

In the farfield, the overlapping light originating from different quantum paths will result in QPI. As light propagates from the focus to the far-field detector, the direction of the light is given by the local gradient of the wavefront. This means that light originating from regions within the focus with the same gradient will overlap in the farfield. This results in interference, where all of the short, long and longer trajectory groups can contribute and interfere with each other (inter-group interference). Due to the Gaussian shape of the intensity focal profile, interference can occur even within one trajectory group (intra-group interference) such as for the long-long interference which can be observed at larger divergences in fig. \ref{fig:macroscopic_interference}, \cite{CarlstromNJP2016}. By varying the peak intensity in the focus, the phase of the photons from the different trajectory groups will vary relative to each-other, and the interference pattern detected in the farfield will change. Recording these QPI patterns as a function of intensity we can extract additional information about the HHG process.

Experimentally, the recorded HHG spectrum results not only from a single half-cycle within the temporal envelope of the driving laser pulse, but from many. As the pulse propagates through the target gas, XUV originating from different half-cycles within the envelope will be generated by different intensities. The result leads to the well known interference effects producing a spectrum of just odd order harmonics for a sinusoidal driving field. However, this means that the QPI resulting from different times throughout the envelope will also differ. The recorded QPI patterns are thus a sum of the generated QPI patterns from all the half-cycles, which we need to take into account if we want to model the result.

To understand the information in the measured QPI plots we compare the experimental results with a simple one dimensional Gaussian beam model. To include the effect of the changing envelope of the driving laser field, the model is sampled at many different times $t$ across the pulse. The generated XUV field $E_{I_0,r}$ is given by following equation 
 \begin{equation}
     E_{I_0,r} = \sum_{t} \sum_{j} A_{I_0,r,t}^n e^{(i(\alpha_{j} I_{I_0, r, t} + \phi_{j}) + i q \omega_{IR} t)},
     \label{eq:2}
 \end{equation}
 
 where the intensity of the IR field can be described by both the temporal and spatial envelopes:
 \begin{equation}
    I_{I_0, r, t} = I_0*e^{-4ln(2)t^2/\tau_{FWHM}^2}e^{-2r^{2}/(r_{0}^{2})}.
    \label{eq:4}
 \end{equation} 
 
 The amplitude of the IR can be expressed as:
 \begin{equation}
    A_{I_0, r, t} \propto \sqrt{I_{I_0, r, t}/I_0},
    \label{eq:5}
 \end{equation} 

Here, $I_0$ is the peak intensity of the driving laser pulse, $\tau_{FWHM}$ is the full width at half maximum (FWHM) driving pulse duration of angular frequency $\omega_{IR}$, $r_0$ is the $1/e^2$ focal radius of the beam, $n$ is the non-linear parameter for the generated harmonics \cite{PerezHernandezOE2009}, and $\phi$ is a trajectory group, $j$, dependent offset phase. The spatial point orthogonal to the laser direction is given by $r$, $q$ defines the harmonic order, and $t$ samples the time once per half cycle of the driving laser pulse. 
To account for the cut-off intensity variation between the different harmonic orders, the amplitude in equation \ref{eq:2} was multiplied by a Sigmoid function,  
  \begin{equation}
     S_{cut-off} = \frac{1}{1+e^{-c_1(I_{I_0, r, t}-I_{cut-off})}}.
     \label{eq:sigmoid}
 \end{equation}
 The constant $c_1$ was adjusted to best fit the experimental results.
 
 In this model, eq. \ref{eq:2}, a Gaussian focal profile for the generated XUV is defined with both an amplitude and a phase component. Comparing the phase part to eq. \ref{eq:1} we see that this follows the same form, except for the phase term arising from sampling at different times, $t$. As the driving pulse amplitude is different for successive sampling times the intensity for each half cycle changes, resulting in different trajectories and different intensity dependent phases of the light. For each time step we assume that there is one fixed intensity applied for each position. This assumption works well for long driving pulses comprising many cycles. To simulate the measured QPI plots, eq. \ref{eq:2} is then propagated to the farfield through a Fourier transform.

 Using this model we are able to turn on and off contributions from different trajectory groups to isolate the observed effects. Sampling of many different half-cycles leads to a reduced contrast of the single half-cycle QPI plot, however the general features remain the same. The simulated QPI result for including just short and long contributions for harmonic 15 of a 1030\,nm driving field is shown in fig. \ref{fig:sim_qpi}\,a). In contrast, fig. \ref{fig:sim_qpi}\,b) also includes contributions from the third trajectory group. In both cases, the contributions from the short trajectories are dominating the structure in the low divergence region. The signal from the short trajectory group has a small maximum divergence due to the much lower alpha value for this group. The direction of the curvature for the short trajectory group interference results from a negative alpha value for this particular harmonic \cite{CarlstromNJP2016}. 
 
  Comparing the interference patterns at large divergence for fig. \ref{fig:sim_qpi}\,a) and b), it is clear that different signatures arise as a result of the different trajectory groups included in the simulation.  The long and longer trajectory groups have higher alpha values, leading to larger maximum divergences. The larger alpha value also results in multiple radians worth of phase difference, producing interference rings in the farfield, and  fringes along the divergence axis in the frequency specific QPI plot. For the case of emission from just the long trajectory group fig. \ref{fig:sim_qpi}\,a) we see that the interference fringes change linearly with the peak intensity, creating distinctive linear striped signature in the QPI plot. When longer trajectory groups are included as well, fig. \ref{fig:sim_qpi}\,b), the QPI plot signature changes into a curved pattern, which arises from the inter group interference. The precise form of this curved pattern is highly dependent on the number of trajectory groups included, the alpha values used for each of them as well as the phase offset used, their relative strengths, and the intensity range scanned. With so many parameters in combination with experimental uncertainties, this makes it very difficult to accurately fit the complicated QPI plots. It should be noted that the probability of longer trajectories recombining having passed the core diminishes rapidly with the number of passes. We thus only consider the third and the fourth trajectory groups here in addition to the short and long in our theoretical calculations.

\section{Measurement and Analysis}

\begin{figure}
    \centering
    \includegraphics[width=0.45\textwidth]{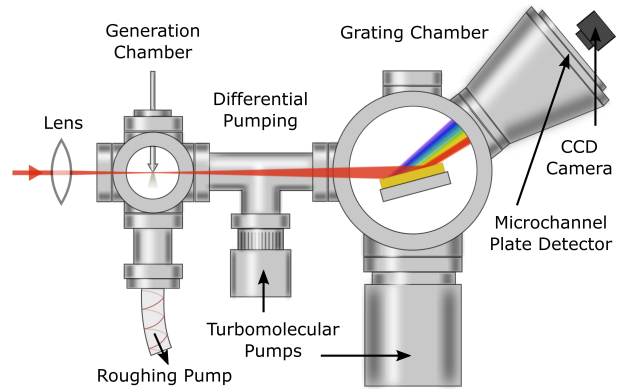}
    \caption{Schematic overview of the HHG setup, consisting of a series of differentially pumped vacuum segments.}
    \label{fig:setup}
\end{figure}

In the experiment we measure and analyse the spectrally resolved intensity dependent interference of different trajectory groups contributing to the HHG process. Our experimental setup comprises a number of interconnected vacuum chambers, represented with a schematic in fig. \ref{fig:setup}.  The harmonic generation process is driven by a Light Conversion PHAROS laser system running at 10\,kHz, delivering $\sim$ 170\,fs FWHM Fourier transform limited pulses around 1030\,nm central wavelength. After estimating propagation losses from the laser output to the interaction region, the pulse energy at the target is approximately 0.6\,mJ. The pulses are focused, with a $f=$10\,cm lens, into a 50\,$\mathrm{\mu}$m aperture continuous gas target which is positioned inside a compact (approximately 1\,dm$^3$) vacuum pumped generation chamber. The peak intensity was estimated to be $1\cdot 10^{18}$\,W/m${}^2$ in the experiment based on the observed spectral cut-off. A target gas of argon was used for the presented results with $\sim$\,3\,bar of backing pressure. The generated harmonics are spectrally separated using a flat-field XUV grating (Hitachi 001-0639) with 600\, lines per mm, and projected onto a microchannel plate detector with a phosphor screen, imaged by a CCD camera.   The generation chamber and the grating chamber are separated by a differential pumping segment, ensuring that the base pressure in the grating chamber can be maintained at sufficiently low vacuum, while the generation chamber is pumped only by a roughing pump.

\begin{figure*}
    \centering
    \includegraphics{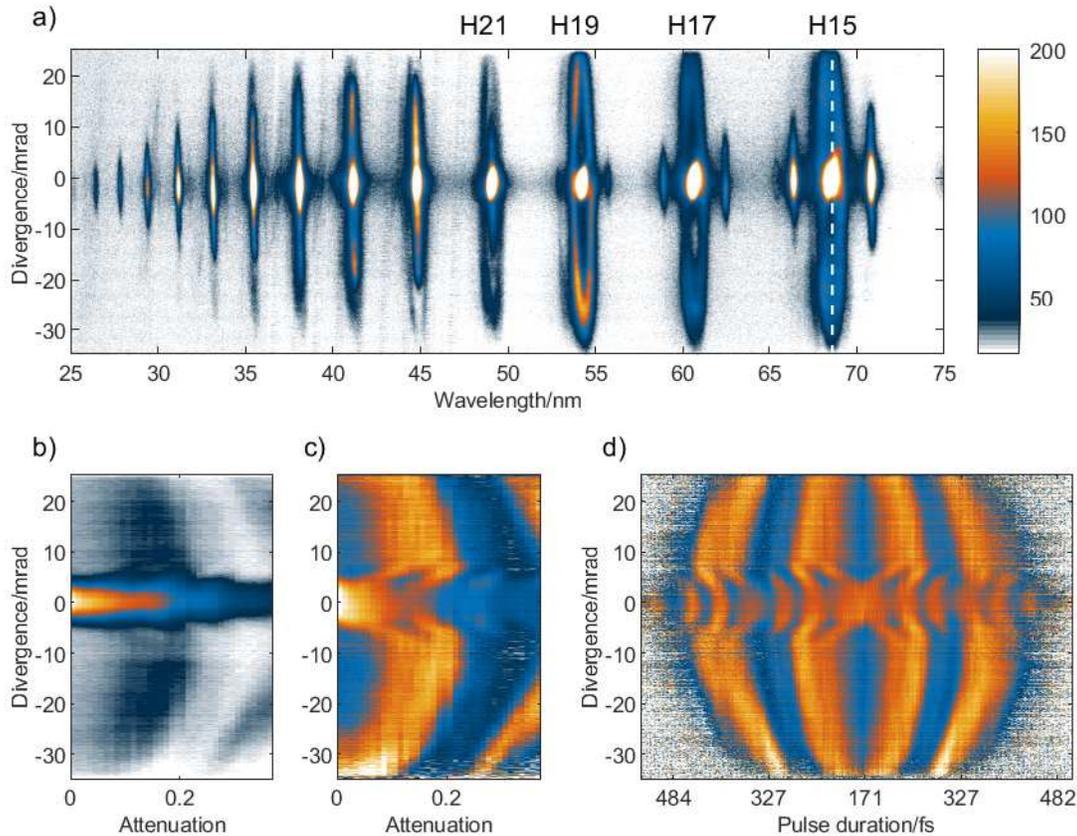}
    \caption{a) Spectrum as recorded on the microchannel plate detector, with the lower order harmonics labelled. b) Extracting the divergence for the center frequency of H15 ($\lambda = 68.6$\,nm) while varying the peak intensity via attenuation. c) Filtering and normalizing the QPI plot in b) to increase the visibility of the interference. d) A filtered QPI plot where the peak intensity is varied by changing the pulse duration through chirping.}
    \label{fig:filtering}
\end{figure*}

 By recording the generated harmonic spectra for each different peak intensity of the laser field we can extract QPI plots showing the variation of the interference pattern as a function of the peak intensity. In fig. \ref{fig:filtering}\,a), we present a typical recorded harmonic spectrum. In each harmonic there is a bright central spot with low divergence indicating the extent of the short trajectory contribution. Longer trajectories are more divergent due to their larger alpha parameters, and thus extend further from the harmonic center. By selecting the central frequency pixel array in each harmonic order for each peak intensity, we are able to build up the QPI pictures. In fig. \ref{fig:filtering}\,b) this is done for harmonic 15 [see the dashed white line in fig. \ref{fig:filtering}\,a)], where the peak intensity of the pulse is varied through attenuation of the laser pulse energy. We observe intensity dependent interference effects, which can be separated into areas where the short trajectories contribute, or do not contribute. To observe the interference effects more clearly, the data is processed using a 2D convolution with a Gaussian function. By taking the ratio between the convolution and the original data, the filtered interference information can be extracted, see fig. \ref{fig:filtering}\,c). This minimizes the prominent background components and increases the visibility of the modulations.
 
 \begin{figure}[!ht]
    \centering
    \includegraphics[width=0.4\textwidth]{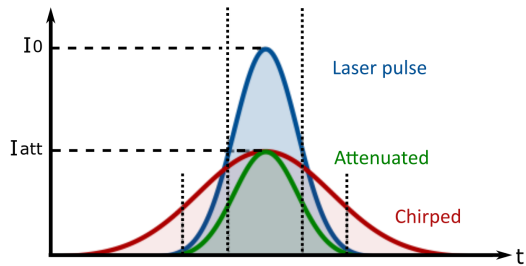}
    \caption{Illustration highlighting the key difference in laser pulse temporal profile when reducing the peak intensity of the full Fourier transform limited pulse shown in blue. The red curve shows intensity reduction through chirping the laser pulse, compared with the green curve where the peak intensity is attenuated directly. }
    \label{fig:cva}
\end{figure}

\begin{figure*}[!ht] 
    \centering
    \includegraphics{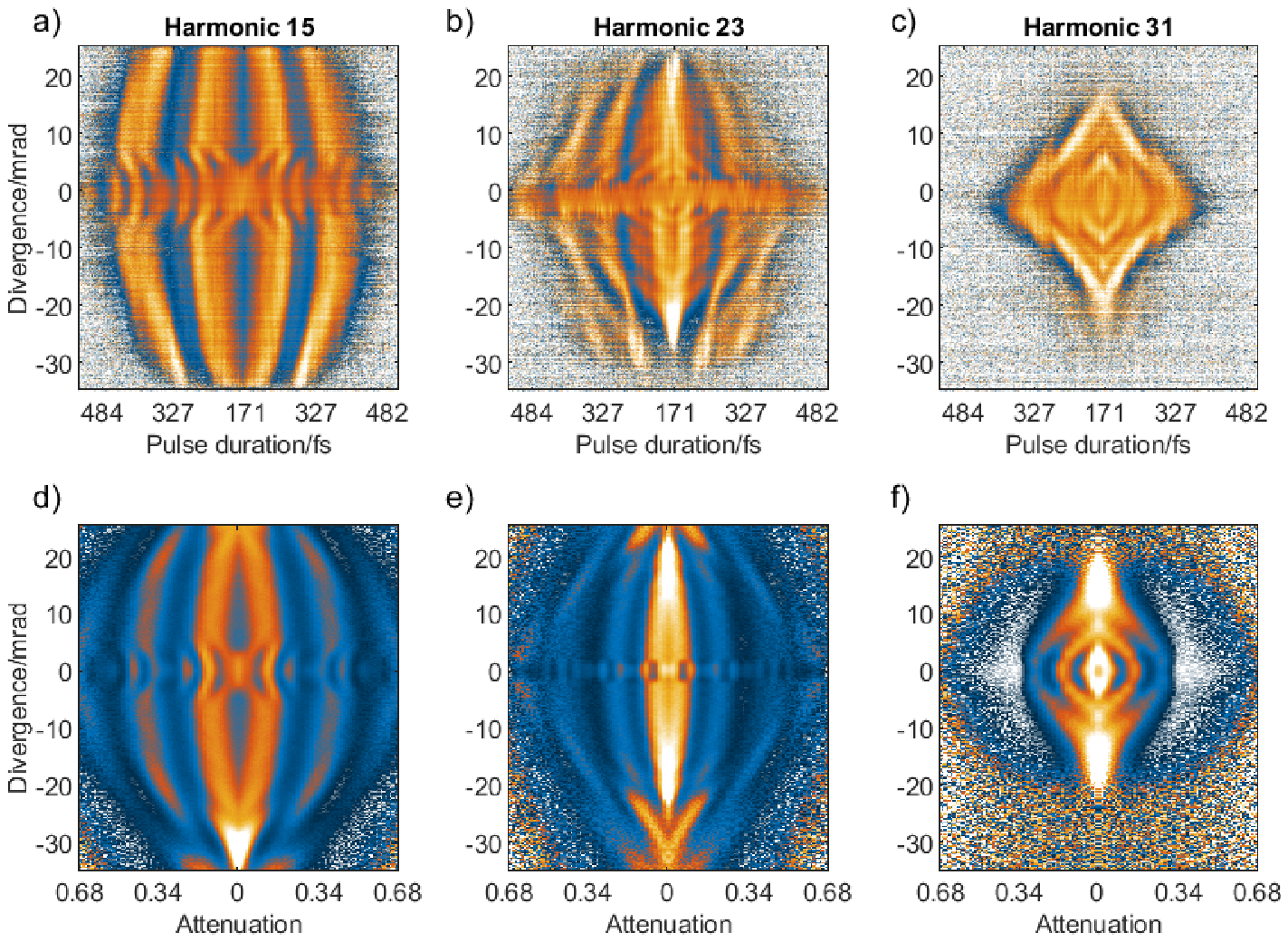}
    \caption{a)-c) Filtered experimental QPI plots for harmonics 15, 23 and 31 respectively, for a driving laser central wavelength of 1030\,nm. The difference between a) and c) highlights the signatures of different contributing trajectory groups, as shown in fig. \ref{fig:sim_qpi}. Comparable simulations using the simple Gaussian model are shown in d)-f). Noise was added before filtering to reduce numerical artifacts. The values used for the fitting can be found in table \ref{table:1}.}
    \label{fig:exp_qpi}
\end{figure*}

 We used two different methods to control the peak intensity of the laser pulse: direct attenuation of the pulse energy, and controlling the chirp of the laser pulse (fig. \ref{fig:cva}). Both of these parameters were adjusted through computer control of the PHAROS laser. Each method has its advantages. Direct attenuation as a method preserves the temporal structure of the pulse and is thus a convenient method for controlling the peak intensity. By performing a chirp scan, however, the pulse is instead stretched in time with a temporal separation of the different frequency components. In this way, the total energy per pulse is preserved while the peak intensity is reduced. To know whether the chirp method can access the same information as the attenuation method we need to determine whether the chirp introduces additional effects in the HHG process other than a variation of the peak intensity. We find that for our experimental conditions the QPIs for the central harmonic frequencies appear symmetric with positive or negative chirp (fig. \ref{fig:filtering}\,d). This indicates an attenuation-like result for that particular frequency. Furthermore, automated control of the laser compressor position allowed us to perform finely resolved chirp scans and this was therefore the main scanning method used.

With the methods described above we observe a stark difference in the QPI pattern for low harmonic orders and high harmonic orders, which we ascribe to different trajectory groups contributing at the different harmonic frequencies.

In fig. \ref{fig:exp_qpi}\,a)-c) we present filtered QPI plots extracted from the chirp scan method showing three different harmonic orders: H15, H23, and H31, as well as simulated QPI plots for these harmonics, fig. \ref{fig:exp_qpi}\,d)-f) (see appendix for how the fitting was done). 
These selected harmonics are chosen to represent different energy regions within the harmonic spectral profile. Harmonic 15 is located deep within the plateau for all trajectory groups. This means that all trajectories can contribute in fig. \ref{fig:exp_qpi}\,a). Near on-axis, where the short trajectories contribute, convex interference fringes are observed which correspond to negative alpha values for low order harmonics, as reported in Carlstr{\"o}m \textit{et al}. \cite{CarlstromNJP2016}. Furthermore, we observe the same signatures when the peak intensity is varied as in fig. \ref{fig:sim_qpi}\,b), where large, concave parentheses shaped interference rings are present off-axis. The strong contrast for this interference between the trajectory groups contributing off-axis indicates that the XUV originating from at least two of these trajectory groups have similar relative amplitudes. We have found it necessary with our model to have strong long and third trajectory contributions to reproduce the distinctive double parenthesis signature in fig. \ref{fig:exp_qpi}\,a). We could also expect even longer trajectory group, for which H15 has a much higher alpha value (see fig. \ref{fig:phase_intensity}\,a), to show up as faster parenthesis frequencies. The fitting parameters used for fig. \ref{fig:exp_qpi}\,d) are presented in table \ref{table:1}. 

    \begin{table*} [!ht]
    \centering
    \begin{tabular}{ l c c c c c  c c c  c c c c  c c c  c c c c }
    & & & &  H 15  & & & & & & & H 23 & & & & & & & H 31\\ 
    [0.5 ex]
    \midrule
    Trajectory Group  & & Short & Long & Third & Fourth & & &   & Short & Long & Third & Fourth & & &   & Short & Long & Third & Fourth \\ 
    [0.5 ex]
    \midrule
    $\alpha$-Value  [$10^{-17}$\,Rad$\cdot$m${}^2$/W] & &-0.4 & 7.7 & 5.4 & - & & & & 0 & 7 & 6.5 & 10.5 & & & & 1 & 6 & - & -  \\[0.5 ex]
    Strength & & 1 & 2 & 2.5 & 0 & & & & 1 & 3 & 2 & 2 & & & & 1 & 2 & 0 & 0 \\[0.5 ex]
    Cutoff Intensity  [$10^{17}$W/m${}^2$] &  & 1 & 1 & 2 & - & & & & 3 & 3 & 4 & 4 & & & & 6.5 & 6.5 & - & -  \\[0.5 ex]
    Nonlinear Parameter & & 4 & 4 & 4 & - & & & & 4 & 4 & 4 & 4 & & & & 4.5 & 4.5 & - & - \\[1 ex]
    \bottomrule
    \end{tabular}
    \caption{Parameter values used for the fitting in fig. \ref{fig:sim_qpi} and \ref{fig:exp_qpi}\,d)-f). The initial values were based on those presented in Carlstr{\"o}m {\textit{et al.}} \cite{CarlstromNJP2016}, but were adjusted to more closely match the divergence of the experimental data as well as other prominent features. The estimated experimental intensity of $1\cdot 10^{18}$\,W/m${}^2$ was used in the theoretical model.}
    \label {table:1}
    \end{table*}

 In the QPI plot for harmonic 31, fig. \ref{fig:exp_qpi}\,c), the observed pattern is clearly different from that for harmonic 15. In this pattern you see the signature of long-long interference, the linear stripes (see fig. \ref{fig:sim_qpi}\,a)) which result in a rhombus shape for the positive and negative symmetrical chirp scan. This difference comes from the harmonic being close to the cut-off region for the short and long trajectories, and above the classical cut-off for the third and fourth trajectory groups. Thus the dominating trajectory groups contributing to this harmonic are the short and the long. 
 
 Although the model can be used to give us a qualitative understanding of the QPI plots, the simple phase model (eq. \ref{eq:1}) becomes increasingly less valid as we approach the cut-off regions, where the difference between the classical approximation and the true quantum mechanical nature of the process is most pronounced. The linear approximation for the intensity dependent phase is thus expected to not fully capture the whole picture around the cut-off \cite{GuoJPBAMOP2018}.
 Within the H31 QPI plot, we observe small deviations from a pure rhombus signature. This could be a result of short-long interference as the alpha values for each of these trajectory groups approach the same value towards the cut-off.  Alternatively, it could be a result of some residual 3rd and 4th trajectory contribution, or contributions from higher order trajectory groups. The fitting parameters used for fig. \ref{fig:exp_qpi}\,f) can be found in table \ref{table:1}. The parameters were chosen in a similar way as for harmonic 15 (see Appendix).

In between these extremes, harmonic 23 is more complex, as the third and fourth trajectory groups are in their cut-off region (fig. \ref{fig:sim_qpi}\,b). The pattern contains contributions of each signature, both the intra-group interference (rhombus shape) as well as the inter-group interference (the parenthesis). The fitting parameters used for fig. \ref{fig:exp_qpi}\,e) can be found in table \ref{table:1}.

Although these signatures can be observed clearly, the details in each of the figures depend on a large number of parameters. Even with our simple Gaussian model, many different considerations must be taken into account in order to qualitatively simulate the QPI data. This includes the alpha values for the four different trajectory groups considered, as well as their relative strengths and initial phase offsets. One step that is unclear is how to relate the pulse duration to attenuation. Experimentally we find that the pulse duration seems to be linear with peak intensity. Looking at the on-axis fringes from harmonic 15, the modulation does not follow the inverse intensity relation that would be expected. Experimental conditions could be the reason for this, for example if we do not generate throughout the pulse due to changed phase matching from plasma formation. This is indicated experimentally by looking at the pixels around the central frequency, where we observe an asymmetry in the chirp scan. Along the divergence axis there is also uncertainty since it depends on the precise size and waveform of the generated XUV in the generating medium, and we intentionally position our focus offset from the gas-jet to phase-match the long trajectories.

When harmonics are generated that are phase matched for both the short and long trajectory groups, indications of longer than long trajectory interference can be seen directly on the MCP detector. These indications in the lab are twofold: first, a striking difference is observed in the more divergent parts of the plateau harmonics. In our case, fig. \ref{fig:filtering}\,a), this can be seen in harmonics 15-23, with H19 having a particularly bright outer ring for this intensity. If only the long-long inter-group interference was the cause of the divergent XUV light, the expectation would be for a higher degree of similarity across the harmonic spectrum. Instead, the variation for the off-axis emission indicates additional frequency-dependent interference effects (i.e. multiple frequency dependent alpha values). Secondly, when observing the harmonic spectrum as the peak intensity is varied the harmonics can also be seen to vary greatly with respect to each other. The different signatures (parenthesis and linear stripes) described above lead to a difference in the dynamics of the intensity dependent interference rings. For the case of harmonics far below any cut-off region (the common plateau), the highly divergent interference rings are seen to emerge from the on-axis center, expand, and disappear as new rings emerge. Instead, for harmonics situated much higher in frequency, the emerging rings as the intensity is increased expand and do not disappear, leading to the linear stripes signature. One can relate this to how a single Gaussian wavefront with increasing intensity is also increasing the maximum divergence of the far-field interference \cite{OlofssonOptLett2021}.
It should be noted that whether a harmonic is considered to be in the plateau or the cut-off is intrinsically intensity dependent.

\section{Conclusions}
We have shown that it is possible to experimentally observe the influence from the electron trajectories that have passed the core before recombination in the high order harmonic generation process by performing a peak intensity scan and recording the frequency-resolved harmonic spectra. Furthermore, with a simple Gaussian beam model we are able to identify the signatures of only long-long, and long-longer trajectory interference in the QPI plots. The signature for the longer-than-long trajectory groups is predominantly observed in the lower harmonic orders, as the cut-off region for the longer trajectory groups occurs at a lower frequency than the short-long cut-off.
We find that the inclusion of trajectory groups longer than long is the reason why the more divergent parts of the harmonics in the common plateau region behave differently to each other.
Future improvements to the measurement accuracy of the experimental parameters could allow for determining the alpha values and strengths of the third and fourth trajectory groups, and could thus be of interest both to compare with theoretical models as well as for a better understanding of the fundamental harmonic generation process.

\begin{acknowledgments}
This research was partially supported by the Swedish Foundation for Strategic Research, the Crafoord Foundation, the Swedish Research Council, the Wallenberg Centre for Quantum Technology (WACQT) funded by the Knut and Alice Wallenberg Foundation, and the Royal Physiographic Society of Lund.

\end{acknowledgments}

\section*{APPENDIX}
The fitting was performed in the following way: first, the long and third trajectory groups were enabled, and their relative cut-off intensities were fitted by looking at the distribution of signal recorded along the x-axis as compared to the experimental QPI plots. Note that for each pair of trajectory groups, the short-long and the third-fourth, the cut-off intensity is the same. The absolute alpha values were then adjusted to match the divergence, and the relative alpha-values were adjusted to fit the periodicity of the parenthesis interference signatures. To match the interference in the center of the QPI plot the alpha-offset values were adjusted. Inclusion of the fourth trajectory could then be done following the earlier procedure. The relative strengths of the different trajectory groups were set in order to match the data. Finally, the short trajectory group could be added following again the previous procedure. Fine adjustments were made to each of the parameters to improve the overall fit to the QPI plot. The above procedure was repeated for each harmonic order, and as many times as needed to get a good qualitative fit.

\bibliography{Ref_lib}

\end{document}